\documentclass[12pt,a4paper,pctex32]{article}
\usepackage{isolatin1}
\usepackage{epsfig}
\def\lsim{\mathrel{\vcenter{\hbox{$<$}\nointerlineskip\hbox{$\sim$}}}}
\def\gsim{\mathrel{\vcenter{\hbox{$>$}\nointerlineskip\hbox{$\sim$}}}}
\newcommand{\nc}{\newcommand}
\nc{\figcap}[1]{\begin{quote}\refstepcounter{figure}
        {\bf Figure \thefigure}: {\small #1}\end{quote}}
\def\mpp{\; \raise1.0ex\hbox{${\scriptstyle +}$\kern-0.85em
      \raise-1.0ex\hbox{${\scriptstyle (-)}$}}\; }
\def\pmp{\; \raise1.0ex\hbox{${\scriptstyle -}$\kern-0.85em
      \raise-1.0ex\hbox{${\scriptstyle (+)}$}}\; }

\usepackage{latexsym}

\nc{\be}[1]{\begin{equation}\mbox{$\label{#1}$}}
\nc{\bea}[1]{\begin{eqnarray} \mbox{$\label{#1}$}}
\nc{\Section}[2]{\section{#2}\label{#1}}
\nc{\Bibitem}[1]{\bibitem{#1}}
\nc{\Label}[1]{\label{#1}}

\nc{\eea}{\end{eqnarray}}
\nc{\ee}{\end{equation}}

\nc{\bdm}{\begin{displaymath}}
\nc{\edm}{\end{displaymath}}
\nc{\dpsty}{\displaystyle}
\nc{\bc}{\begin{center}}
\nc{\ec}{\end{center}}
\nc{\ba}{\begin{array}}
\nc{\ea}{\end{array}}
\nc{\bab}{\begin{abstract}}
\nc{\eab}{\end{abstract}}
\nc{\btab}{\begin{tabular}}
\nc{\etab}{\end{tabular}}
\nc{\bit}{\begin{itemize}}
\nc{\eit}{\end{itemize}}
\nc{\ben}{\begin{enumerate}}
\nc{\een}{\end{enumerate}}
\nc{\bfig}{\begin{figure}}
\nc{\efig}{\end{figure}}

\nc{\arreq}{&\!=\!&}
\nc{\arrmi}{&\!-\!&}
\nc{\arrpl}{&\!+\!&}
\nc{\arrap}{&\!\!\!\approx\!\!\!&}
\nc{\non}{\nonumber}
\nc{\align}{\!\!\!\!\!\!\!\!&&}

\def\lsim{\; \raise0.3ex\hbox{$<$\kern-0.75em
      \raise-1.1ex\hbox{$\sim$}}\; }
\def\gsim{\; \raise0.3ex\hbox{$>$\kern-0.75em
      \raise-1.1ex\hbox{$\sim$}}\; }

\nc{\DOT}{\hspace{-0.08in}{\bf .}\hspace{0.1in}}
\nc{\Laada}{\hbox {$\sqcap$ \kern -1em $\sqcup$}}
\nc\loota{{\scriptstyle\sqcap\kern-0.55em\hbox{$\scriptstyle\sqcup$}}}
\nc\Loota{{\sqcap\kern-0.65em\hbox{$\sqcup$}}}
\nc\laada{\Loota}
\nc{\qed}{\hskip 3em \hbox{\BOX} \vskip 2ex}

\nc{\real}{{\rm I \! R}}
\nc{\Z}{{\sf Z \!\!\! Z}}
\nc{\complex}{{\rm C\!\!\! {\sf I}\,\,}}
\def\bigid{\leavevmode\hbox{\small1\kern-3.8pt\normalsize1}}
\def\id{\leavevmode\hbox{\small1\kern-3.3pt\normalsize1}}
\nc{\slask}{\!\!\!/}
\nc{\bis}{{\prime\prime}}
\nc{\pa}{\partial}
\nc{\na}{\nabla}
\nc{\ra}{\rangle}
\nc{\la}{\langle}
\nc{\goto}{\rightarrow}
\nc{\swap}{\leftrightarrow}

\nc{\EE}[1]{ \mbox{$\cdot10^{#1}$} }
\nc{\abs}[1]{\left|#1\right|}
\nc{\at}[2]{\left.#1\right|_{#2}}
\nc{\norm}[1]{\|#1\|}
\nc{\abscut}[2]{\Abs{#1}_{\scriptscriptstyle#2}}
\nc{\vek}[1]{{\rm\bf #1}}
\nc{\integral}[2]{\int\limits_{#1}^{#2}}
\nc{\inv}[1]{\frac{1}{#1}}
\nc{\dd}[2]{{{\partial #1}\over{\partial #2}}}
\nc{\ddd}[2]{{{{\partial}^2 #1}\over{\partial {#2}^2}}}
\nc{\dddd}[3]{{{{\partial}^2 #1}\over
    {\partial #2 \partial #3}}}
\nc{\dder}[2]{{{d #1}\over{d #2}}}
\nc{\ddder}[2]{{{d^2 #1}\over{d {#2}^2}}}
\nc{\dddder}[3]{{d^2 #1}\over
    {d #2 d #3}}
\nc{\dx}[1]{d\,^{#1}x}
\nc{\dy}[1]{d\,^{#1}y}
\nc{\dz}[1]{d\,^{#1}z}
\nc{\dl}[1]{\frac{d\,^{#1}l}{(2\pi)^{#1}}}
\nc{\dk}[1]{\frac{d\,^{#1}k}{(2\pi)^{#1}}}
\nc{\dq}[1]{\frac{d\,^{#1}q}{(2\pi)^{#1}}}

\nc{\bfT}{{\bf T }}

\nc{\cA}{{\cal A}}
\nc{\cB}{{\cal B}}
\nc{\cD}{{\cal D}}
\nc{\cE}{{\cal E}}
\nc{\cG}{{\cal G}}
\nc{\cH}{{\cal H}}
\nc{\cL}{{\cal L}}
\nc{\cO}{{\cal O}}
\nc{\cT}{{\cal T}}
\nc{\cN}{{\cal N}}
\nc{\cR}{{\cal R}}
\nc{\spa}{\partial\!\!\!\!\,\raise0.1ex\hbox{/}}
\nc{\rvac}[1]{|{\cal O}#1\rangle}
\nc{\lvac}[1]{\langle{\cal O}#1|}
\nc{\rvacb}[1]{|{\cal O}_\beta #1\rangle}
\nc{\lvacb}[1]{\langle{\cal O}_\beta #1 |}
\nc{\bb}{\bar{\beta}}
\nc{\bt}{\tilde{\beta}}
\nc{\ctH}{\tilde{\cal H}}
\nc{\chH}{\hat{\cal H}}
\nc{\1}{\aa}
\nc{\2}{\"{a}}
\nc{\3}{\"{o}}
\nc{\4}{\AA}
\nc{\5}{\"{A}}
\nc{\6}{\"{O}}
\nc{\al}{\alpha}
\nc{\g}{\gamma}
\nc{\Del}{\Delta}
\nc{\e}{\textrm{e}}
\nc{\eps}{\epsilon}
\nc{\lam}{\lambda}
\nc{\Om}{\Omega}
\nc{\ve}{\varepsilon}
\nc{\mn}{{\mu\nu}}
\nc{\vp}{\varphi}
\begin{document}
\baselineskip=20 pt

\title{Singlet fermions on curved extra dimension tori}
\author{ I. Vilja\thanks{vilja@utu.fi} \\
\emph{Department of Physics, FIN-20014 University of Turku,
                                              Finland}}
\maketitle

\begin{abstract}
A model with two curved compact extra dimensions is introduced.
The model is based on a four-brane immersed in a six dimensional 
space, where the extra dimensions are compact but not flat. They 
have topology of torus. The form of metric in the empty bulk is 
studied and the gauge singlet fermion structure showed to be very simple. 
It contain only one massless low energy mode which couples to brane 
matter, while the massive modes are not related to the volume of torus 
while the Planck mass is related to the volume of extra dimensions. In the 
model "our" brane is situated on a nearly singular line of the torus.
\end{abstract}

\vspace{-14cm}

\thispagestyle{empty}
\newpage

\section{Introduction}

The intriguing idea, that the physical space may have more that three dimensions
has got a lot attention recently. The idea itself dates bect to the beginning of the
20th century, to the works of Nordst\" om, Kaluza and Klein \cite{0}. They
attempted to
unify gravity and electromagnetism by introducing a fifth dimension. The idea
reappeared in string theories, where unification all four known fundamental 
interactions  was made in ten or eleven dimensional space-time. Originally the
string theorists supposed, that the left-over dimensions are compactified to 
extremely short distances, which therefore would have only a little (direct) impact to
low energy physics and particle phenomenology. Later, however, it was found that
these extra dimensions, or some of them, can be relatively large, lie on an 
intermediate
scale \cite{0+}. This finding is related to the uncertainty of  the measurements of 
gravitational forces. They indicate, that the extra dimensions may be even as large as
 $\sim 1$ mm \cite{4,5, 5b}.
These considerations have boosted a new wave of research on large extra
dimensions because the may have testable consequences in present day or near
future experiments.

In the core of this research is the notion, that the Planck scale $M_{Pl}$ is possibly 
not the "true" gravity scale, but only an effective gravity scale related to
our four dimensional space-time $M_4$ of our natural perception \cite{1,2,3}.
The gravity scale $M_*$  of the space-time which includes $M_4$ and $n$ the 
large extra dimensions, is related to Planck scale by a well defined equation
$M_{Pl}\sim M_*^{n+2}V^{(n)}$, where $V^{(n)}$ is an extra dimension
 size related factor. Depending on the form of   $V^{(n)}$, $M_*$ may be considerably 
lower scale than $M_{Pl}$. 
Moreover, the particles of the Standard model (SM) or other gauge symmetries  are 
supposed to live only on a four dimensional slices of all large dimensions 
({\it e.g.} on $M_4$) \cite{2}. This slices, called the branes, may have at least 
two possible realizations. They may be defined by
solitonic configurations of underlying string theory, when they have finite but
small width being effectively four dimensional. Or they may be constructed as a singular 
point of an orbifold, when they are real four dimensional 
manifolds \cite{7b}.

In these large extra dimension models the particles which are gauge singlets
may propagate outside the brane(s), in so called bulk \cite{2}, forming bulk
matter.
Interactions between the brane and bulk matter may therefore rise interesting
consequences. Righthanded neutrinos may be this kind singlet fermions \cite{6,7}.
This would explain quite naturally {\it e.g.} the lightness of neutrinos: their masses
would be suppressed by factor $\sim M_*/M_{Pl}$. A lot of work on particle
phenomenology of extra dimension models has done during last years \cite{7c,8}. 
Also the cosmological implications of the extra dimensions have been subject 
of numerous studies \cite{7b, 7d}. 

In practise, there is two fundamental types of extra dimension constructions.
One may have compact extra dimensions, so called ADD-type models \cite{1}, or
the extra dimensions may be non-compact but "wrapped", when one speaks of
RS-models \cite{8b}. In ADD-models the factor  $V^{(n)}$ is simply the volume
of the extra dimensions, while it in RS-models is merely related to the 
normalization of gravitons. In either case the particles living in the bulk shows up to 
form to an infinite Kaluza-Klein tower of excitations when observed on the brane.

In the present paper we study a model which lies somewhere in-between
ADD-models and RS-models. The manifold of the extra dimensions is a compact one, 
and the bulk empty. We study a four-brane embedded on a six dimensional space of
large dimensions. However we do not suppose it to be a flat manifold but allow it be 
curved, somehow wrapped like in RS-models. This makes possible to obtain a 
considerablysimple spectrum of excitations of the bulk matter, in particular bulk 
fermions as the right handed neutrinos.  We also  study the general features
of this model and propose a more detailed, simple construction.

\section{Geometry on a torus}

We coordinatise the six dimensional space with $z^M$, where $M=0,\dots , 5$,
where
the first four coordinates are associated to the Minkowski space.
The two extra dimensions have topology of two-torus T$^2$ with coordinates
$y^a$, $a=1,2$. The metric of these extra dimensions,
$ds_E^2 = g_{ab}dy^a dy^b$, can quite generally be written in the form 
($y^1=\phi,\ y^2=\theta$)
\be{eq1}
ds_E^2 = a^2 d\phi^2  + f(\phi, \theta)^2 d\theta^2,
\ee
applying suitable coordinate transformations. Here we can assume, that $a$ is a constant and
the range of the angle variables $\phi$ and $\theta$ is the full circle. 
Because of a stationary 
situation, the metric of extra dimensions do not depend on Minkowski space coordinates, in 
particular on time: Riemann and Ricci curvature tensors do not mix Minkowski and 
extra dimension components. Thus the only non trivial components are 
$R_{\theta\phi\theta\phi} = -ff_{\phi\phi},\ R_{\theta\theta}=ff_{\phi\phi}/a^2$ and
$R_{\phi\phi}=f_{\phi\phi}/f$ and hence the curvature scalar of the extra
dimensions reads ${\cal R} = - 2 f_{\phi\phi}/(a^2 f)$. The volume of the torus can be
also calculated, giving naturally 
\be{vol}
V_2= a \int_0^{2\pi}d\theta \int_0^{2\pi}d\phi f(\phi,\theta ).
\ee
The Einstein tensor $E^\mu_\nu = R^\mu_\nu - \frac 12 g^\mu_\nu \cR$ reads now 
$E^\mu_\nu =  - \delta^\mu_\nu \cR$ while extra dimension components are
equal to zero:  $E^a_b=E^a_\mu=0$. 
The six dimensional Einstein equation
\be{Ee}
E^A_B= 8\pi G_6 T^A_B
\ee
implies thus, that the bulk is empty and possible branes are located in
$\phi$-direction at points $\phi_i$. 
Eq. (\ref{Ee}) means that one has indeed a four branes instead of a three-branes:
the position of brane is not fixed in $\theta$-dimension, but includes it as whole. 
Thus the geometry describes empty bulk with branes of tension $\sigma_i$ located given 
points. The piecewise defined bulk solution, the scale function $f$ consists of simple 
positive linear pieces
\be{solf}
f(\phi , \theta )= f_i(\theta ) + [f_{i+1}(\theta ) -  f_i(\theta )]
{\phi-\phi_i\over \phi_{i+1}-\phi_i},\ \ \phi_i \le \phi < \phi_{i+1},
\ee
where, without loss of generality can be chosen $\phi_1=0,\ \phi_{K+1}=2 \pi$ 
representing the same point. Note, that due to periodicity the number 
of branes is $0$ or larger than $1$. Naturally,
if it equals zero, $f$ is a function  of $\theta$ only and this function can be 
absorbed to
the coordinate itself.  The positions of branes, functions $f_i$ and the brane tensions
$\sigma_i$ are related by crossing rules, the jump conditions, which reads as
\be{jump}
{\Delta f_i \over \Delta \phi_i} - {\Delta f_{i-1} \over \Delta \phi_{i-1}}=
-4\pi M_{*}^{-4} \sigma_i a^2 f_i,
\ee
where $\Delta f_i = f_{i+1}-f_i$ and $\Delta \phi_i = \phi_{i+1} -\phi_i$ and 
$M_*$ is the six-dimensional fundamental gravity scale related to six-dimensional
gravity  constant by $G_6=M_*^{-4}$.  With this notation
The volume of extra dimensions reads
\be{vol2}
V_2= a \int_0^{2 \pi}d\theta\sum_i \Delta \phi_i {f_{i+1}+f_i\over 2},
\ee
and relation between the four-dimensional gravitational scale 
(the Planck scale $M_{Pl}$) and
six-dimensional scale $M_*$ reads now as 
$M^2_{Pl}= M_*^4 \int d^2y\, \sqrt {-g_6} =V_2 M_*^4$.
Of course, if functions $f_i$ are independent on  coordinate $\theta$, the integration 
contributes only with factor $2\pi$. Moreover, if there are also only two branes,
the jump conditions are simply,
\be{jump2}
\Delta f \left ( \frac 1{2\pi - \phi'}+\frac 1{\phi'}\right ) = - 16\pi\, G_6\, a^2
 \sigma_1 f_1 = 16\pi\, G_6\, a^2 \sigma_2 f_2, 
\ee 
and the volume reads
\be{vol3}
V_2= 2\pi^2 a\, (f_1+f_2), 
\ee
where $\Delta f = f_2 -f_1$, and $\phi' \ne 0$ is the position of the other brane. Should be again
emphasized, that one single brane is not possible because it leads to constant $f$ at
whole bulk: it implies no discontinuities, no tension and no matter. 

Because  our aim is to calculate the equations of motion for fermion fields, we need the spin 
connection defined by
\be{sk}
\Omega_M = \frac 12 \Gamma^{AB}V_A^N\partial_MV_{BN},
\ee
where $V_A^N$ is the moving frame (sixbein) $V_A^N = {\rm diag}(1,1,1,1,a,f)$,
$\Gamma^{AB} = \frac i2 [\Gamma^A,\Gamma^B]$ and matrices $\Gamma^A$ are
the $8\times 8$ gamma matrices for six dimensional space. In this presentation we use the
gamma matrices
\be{ga}
\Gamma^\mu =\left( \begin{array}{cc} 0 & \gamma^\mu
      \\ \gamma^\mu & 0 \end{array} \right), \ \
\Gamma^4 =\left( \begin{array}{cc} 0 & 1_4
      \\ -1_4& 0 \end{array} \right),\ \
\Gamma^5 =-i \left( \begin{array}{cc} 0 & \gamma^5
      \\ \gamma^5 & 0 \end{array} \right)\ ,
\ee
where $\gamma^\mu$'s are the usual four-dimensional gamma matrices. For
the 2-torus metric (\ref{eq1}), however,  all components of the spin connection vanish.
Thus the covariant derivative for spin $\frac 12$-field reads simply 
${\cal  D}_A = V_A^M\partial_M$. Note also, that the 
six-dimensional chirality operators are now given by $P_\pm =\frac{1}{2} (I_8 \pm
\Gamma^7)$, where $\Gamma^7 =- \Gamma^0 \Gamma^1 \cdots \Gamma^5
=\mbox{diag}(1_4 ,-1_4)$ is the generalization of $\gamma^5$ in six dimensions.

If the dimension coordinated by $\theta$ is a maximally symmetric subspace, one can even 
choose a coordinate system where $f$ in independent of $\theta$-coordinate itself.  It is
naturally equivalent with the assumption, that the function $f$ can be split to product of
separate $\theta$ and $\phi$ dependent parts.

\section{Singlet fermions on a general torus}

In particle theory models the gauge-singlet fermions, which are thus not confined on a
brane, live on the whole six-dimensional space. The most natural such particles are the 
singlet neutrinos, which may appear even in the simplest extensions of the standard model.
To study their properties on the manifold $M_4\times T^2$ we have to write down the 
action for the singlet neutrinos. The free field Lagrangian reads as
\be{A6}
S_N =\int d^4x\, d^2 z V [ i\bar N\Gamma^A {\cal  D}_A N + {\rm h.c.}
 - M \bar N N],
\ee
where $V=\det(V_A^N)=a\,f$, $N$ 
is the fermion field at six dimensions, $\bar N = N^\dagger \Gamma^0$ and
$M$ is the mass of the fermion field. We also denote chiral six dimensional eight component
fermions as $N_\pm=P_\pm N$ and identify them with four component chiral fermions, i.e.
\be{Ncomp}
N=
\left (\begin{array}{c} N_+\\N_- \end{array}\right ).
\ee
Note, that $N_\pm$ are identifiable to four dimensional Dirac fermions.
Using these components, the action reads
\bea{A6b}
S_N &=&\int d^4x\, d^2 z V \left [ i(\bar N_-\ \bar N_+)
\left ( \begin{array}{cc} 0&1\\ 1&0\end{array}\right)\otimes \gamma^\mu \partial_\mu 
\left ( \begin{array}{c} N_+\\N_-\end{array}\right )\right .
\nonumber \\
& &+{i\over 2} (\bar N_-\ \bar N_+)
\left ( \begin{array}{cc} 0&1\\ -1&0\end{array}\right)\otimes \id\frac 1a \partial_\phi
\left ( \begin{array}{c} N_+\\N_-\end{array}\right )
+ {\rm h.c.}\nonumber\\
& & +\ \  \frac i2 (\bar N_-\ \bar N_+)
\left ( \begin{array}{cc} 0&\gamma^5\\ \gamma^5&0\end{array}\right)\otimes \id
\frac {-i}f \partial_\theta
\left ( \begin{array}{c} N_+\\N_-\end{array}\right )
+ {\rm h.c.}\nonumber\\
& & \left . +\ \  M  (\bar N_-\ \bar N_+) \left ( \begin{array}{c} N_+\\N_-\end{array}
\right )\right ],
\eea
where now $\bar N_\pm = N_\pm^\dagger \gamma^0$. The corresponding equation 
of motion reads now
\be{eqm}
- i\gamma^\mu \partial_\mu \left ( \begin{array}{c} N_+\\N_-\end{array}\right ) =
\frac ia \partial_\phi \left ( \begin{array}{c} - N_+\\N_-\end{array}\right ) +
i\gamma^5 \frac {-i}f \partial_\theta
\left ( \begin{array}{c} N_+\\N_-\end{array}\right )+
\frac i{2a} {\partial_\phi f\over  f}
\left ( \begin{array}{c} - N_+\\N_-\end{array}\right )+
M  \left ( \begin{array}{c} N_-\\N_+\end{array}\right ).
\ee
This equation can be solved by separating Minkowski space and torus variables leading to
tedious calculations, where the components mix strongly. Therefore,
we henceforth assume that the six-dimensional field $N$ is chiral field in the sense
of six-dimensional theory, e.g. $N_-=0$. This corresponds the assumption $M=0$, which 
decouple the six dimensional chiral fields. It should be also noted, that for chiral 
six-dimensional spinors there are no separate Majorana-couplings either, because 
Majorana fermions at six dimensions are chiral themselves.

Hence we may rewrite the equation of motion for $N_+$ field alone as
\be{eqm+}
- i\gamma^\mu \partial_\mu N_+ = - \frac ia \partial_\phi N_+ -
\frac i{2a} (\partial_\phi \ln f ) N_+
+\frac 1f \gamma^5\partial_\theta N_+ .
\ee
In this equation we perform a generalized Fourier series expansion: The right side of 
eq. (\ref{eqm+})  is set equal to $- M_+N_+$, where
 $M_+ = m_+ + i \mu_+\gamma^5$  and $m_+$ and $\mu_+$ are
real numbers. Then we write the field as a sum of its chiral 
components $ N_+ = N_{++} + N_{+-}$ with
 $ \gamma^5 N_{+\pm} = \pm N_{+\pm}$,
so that the equation of motion for the fields reads as
\bea{eqm+2}
 - \frac ia \partial_\phi N_{++} +\frac 1f \partial_\theta N_{++} - \frac i{2a} 
(\partial_\phi \ln f ) N_{++}
&=& - M_+N_{++},\nonumber\\
- \frac ia \partial_\phi N_{+-} -\frac 1f \partial_\theta N_{+-} - \frac i{2a} 
(\partial_\phi \ln f ) N_{+-}
 &=& - M_+N_{+-}.
\eea

By writing the chiral fields as $N_{+\pm} = H_\pm(\phi, \theta) \psi_\pm(x)$, 
where $\psi_\pm$ are usual four dimensional fermion fields, one finds, that the 
equations of motion for $H_\pm$  are
\be{eqmH}
-\frac ia \partial_\phi H_\pm \pm\frac 1f \partial_\theta H_\pm 
- \frac i{2a} (\partial_\phi \ln f ) H_\pm
= - (m_+ \pm i\mu_+)H_\pm,
\ee 
where upper/lower signs belong to same equation. Immediately one sees, that
\be{st1}
H_\pm(\theta\; ,\phi) = e^{i\nu_\pm \theta}K_\pm (\phi),
\ee
where $\nu_\pm$ are due to $\theta$-periodicity integers and $K_\pm$ are functions 
of $\phi$ only. Inserting (\ref{st1}) to  (\ref{eqmH}), one finds solution
\be{st2}
K_\pm (\phi ) = \exp \left [i\, a \int_0^\phi du \left (- m_+
+ \frac i{2a} \partial_u \ln f(u) 
 \pm i\left (- \mu_+ -
{\nu_\pm\over f(u)}\right ) \right ) \right ] K_\pm (0).
\ee 
Again, consistency with periodicity requires, that $K_\pm (2\pi ) = K_\pm (0)$ and 
we obtain relation for mass parameters $m_+$ and $\mu_+$:
\be{condm}
- a\, m_+ = n,\ \ \ n\in \Z
\ee 
and\be{condmu1}
\int_0^{2\pi }d\phi  \left (\mu_+ +
{\nu_\pm\over f(\phi )}\right ) = 0.
\ee
Note, that $ \int_0^{2\pi} du\, \partial_\phi \ln f (u) = f(2\pi ) - f(0)$ vanish 
automatically due to  periodicity of $f$.
Moreover, Eq. (\ref{condmu1}) should hold for both $\nu_+$ and $\nu_-$.
Thus we find that they have to be
equal $\nu_+ = \nu_- \equiv \nu$ and
\be{condmu}
\mu_+  = - \frac \nu{2 \pi} \int_0^{2\pi }\frac {d\phi}{f(\phi)},\ \ \ \nu \in \Z.
\ee
Thus two integers $n$ and $\nu$ determine the solution expect a constant $K_\pm (0)$
which can be absorbed to four dimensional spinors $\psi_\pm$. 
If we, moreover, define a matrix function
\be{mH}
H_+^{n\nu } = {1\over \sqrt {V_2}}\left( {f(0)\over f(\phi)}\right )^{1/2}
 e^{i \nu \theta + i n \phi} e^{a \int_0^\phi d\phi \left (\mu_+ + {\nu\over f(\phi )}
\right )\gamma^5},
\ee
the solution of general linear equation (\ref{eqm+}), the field $N_+$ can be written as a 
generalized Fourier expansion
\be{solN}
N_+ = \sum_{n, \nu } H_+^{n\nu } \psi^{n\nu},
\ee
where $\psi^{n\nu} = \psi_+^{n\nu} +\psi _-^{n\nu}$ are the corresponding Fourier 
coefficients of each mode $n, \nu$, {\it i.e.} the 
four dimensional fermion fields.

\section{Effective four dimensional singlet fermions}

The next task is to calculate the spectrum of fermions of the four dimensional effective 
model. Indeed there will be a double tower of exited states $\psi^{n\nu}$ when the 
extra dimensions are integrated out. These modes do mix with each other, because the 
functions $H_+^\pm$ are not orthogonal in the space of extra dimensions. Thus 
there appear complicated mixings from both kinetic and mass terms, which should, 
in principle, be diagonalized.

We consider chiral ($M=0,\ N_- =0$) version of Eq. (\ref{A6}):
\be{A7}
S_N=\int d^4x\, d^2z\, V\, [\frac i2 \bar N_+\Gamma^A \cD_A N_+ 
- \frac i2 \cD_A \bar N_+\Gamma^A  N_+],
\ee
which now can be cast in the form involving four dimensional spinors, 
\be{A8}
S_N=\int d^4x\, d\phi\, d\theta\, a f(\phi) \left [  i \bar N_+\gamma^\mu
 \partial_\mu
N_+ -\frac 12\, \bar N_+ M_+ N_+  -\frac 12\, \overline{(M_+ N_+)}N_+\right ],
\ee
where all indices are suppressed.
By inserting the expansion (\ref{solN}) into preceding equation we are able to integrate
over the extra angular coordinates, {\it i.e.} write the action in the form of pure
four dimensional theory. However, this four-dimensional action is quite complicated,
 because neither potential nor kinetic term  are diagonal with respect the index $n$. 
One obtain the action for four-dimensional spinors
\bea{4action}
S &=& \int d^4x \Bigg \{  \sum_{n\, n'\, \mu} \left [
i A^+_{n\, n'\, \nu} \bar \psi_+^{n\, \nu} \spa\psi_+^{n'\, \nu}
+ i A^-_{n\, n'\, \nu} \bar \psi_-^{n\, \nu}\spa\psi_-^{n'\, \nu} \right ]
\nonumber\\
& &\qquad -
\sum_{n\,\nu} B \bar m\, \bar \psi_-^{n\, \nu}\psi_+^{n\, \nu} + {\rm h.c.}
 \Bigg \} ,
\eea
where $\bar m$ stands for $\bar m= m_+ + i\mu_+$ using Eqs. (\ref{condm}) and  
(\ref{condmu}) and, finally, the coefficients of the various terms are
\bea{coeff1}
A^+_{n\, n'\, \nu} &=& {2\pi a\, f(0)\over V_2} \int_0^{2\pi} d\phi\, 
e^{-i a \Delta m\, \phi} e^{+ 2 a \int_0^\phi (\mu_+ + {\nu\over f(u )})du},\non\\
A^-_{n\, n'\, \nu} &=& {2\pi a\, f(0)\over V_2} \int_0^{2\pi} d\phi\, 
e^{-i a \Delta m\, \phi} e^{- 2 a \int_0^\phi (\mu_+ + {\nu\over f(u )})du},\\
B &=& {4\pi^2 a\, f(0)\over V_2}\non
\eea
where  $\Delta m = (n - n')/a$. Note, that $B$ is independent on all indices.

Thus we have obtained a infinite double tower of fermionic excitations
where kinetic mixings are present. However, at this base of functions the mass terms are 
diagonal. Of course, if $f$ is a constant function, as on
flat torus, all non-diagonal terms vanish, because $\mu_+ +\frac \nu f \equiv 0$, and 
modes decouple from each other.

To proceed, one needs a viable approximation which simplifies the mixed up situation.
First, we assume, that there is only two branes. Then a $\theta$-independent scale function 
defined on the interval $\phi \in [0, 2\pi)$ is now
\be {f2}
f(\phi ) = H(\phi' - \phi ) (f_1 + \Delta f \frac \phi{\phi'}) + 
H(\phi - \phi' ) (f_1 + \Delta f \frac {2\pi - \phi}{2\pi - \phi'}),
\ee
where $ \Delta f$ and $\phi'$ are as previously. The jump condition and the volume are
given by Eqs. (\ref{jump2}) and (\ref{vol3}), respectively. From Eq. (\ref{condmu})
one reads now
\be{cmu}
\mu_+=- {\nu\over\Delta f} \ln {f_2\over f_1}=- {\nu\over\bar f},
\ee
where thus $\bar f^{-1} =  {1\over\Delta f} \ln {f_2\over f_1}$ is a kind
effective size of the $\phi$-direction of the torus.

Suppose now, that $f_2\gg f_1$, we find that the volume of extra dimensions
is essentially $V_2=2\pi^2 a f_2$ and the six-dimensional gravity scale reads
 $M_*^4= M_{Pl}^2/(2\pi^2 a f_2)$. We then take into account the physical 
requirement , that $M_* \gsim 10 $ TeV. Also, because we really have now a
four brane, where usual non gauge-singlet particles are  bounded with one compact 
dimension of radius $a$, the massive Kaluza-Klein modes are to be pushed beyond 
observational limit. The lightest  Kaluza-Klein mode has a mass of
$2\pi/a$ which thus has to be high enough, say $2\pi/a \gsim 10$ TeV, too, 
{\it i.e.} $a \lsim 0.6\ {\rm TeV}^{-1}$. Taking the upper limit for $a$ and 
lower limit for $M_*$, and combining these two physical conditions one finds,
that $f_2$ is large,  $f_2\sim 1.2\times 10^{24}  {\rm GeV}^{-1} = 
2.3\times 10^8$ m.  Inserting these values to Eqs. (\ref{coeff1}), we find
that the ratios $A^\pm_{n\, n'\, \nu}/ A^\pm_{n\, n\, \nu}\sim 10^{-16}$ 
( $\phi' \sim \pi$, $n'\ne n$) and $ A^\pm_{n\, n\, \nu}$ in practise
 independent on $\nu$ with  $ A^\pm_{n\, n\, \nu} = B$. Thus, for practical 
purposes the four-dimensional effective spinor action (\ref{4action}) is diagonal.

We may once more write the spinor action using assumptions given above.
The action reads
\be{4actionb}
S = \int d^4x \Bigg \{  \sum_{n \mu} \left [
i \bar \psi_+^{n\, \nu} \spa\psi_+^{n\, \nu}
+ i \bar \psi_-^{n\, \nu}\spa\psi_-^{n\, \nu}
- \bar m_{n\,\nu}\, \bar \psi_-^{n\, \nu}\psi_+^{n\, \nu} + {\rm h.c.}
 \right ] \Bigg \} ,
\ee
where we have scaled fields as 
$\psi_\pm^ {n\, \nu} \to \psi_\pm^ {n\, \nu}/\sqrt B$ and made to each 
mode  a suitable chiral rotation in purpose to
remove complex mass. The mass $\bar m_{n\,\nu}$ reads now
\be{effmass}
\bar m_{n\,\nu} = \sqrt { \left( {n\over a}\right )^2 +
\left( {\nu\over \bar f}\right )^2 }.
\ee
Thus $N\ne 0$ excitations are readily heavy and above direct observational limits.
When one requires that also $n=0,\ \nu \ne 0$ excitations are not directly observable, 
one has to require $\bar f$ to be small enough. Because
$\bar f \simeq f_2/\ln{f_2\over f_1}$,  where $f_2 \sim 10^{24}\ {\rm GeV}^{-1}$, 
one has to push $f_1$ very low, nearly zero. Indeed one sees,
that $\phi=0$ brane lies almost at a gravitational singularity. The geometry of
this torus can be compared to real doughnut submerged to ${\real}^3$: The perimeter 
of the inner ring  is smaller than the perimeter of the outermost ring
while at the perpendicular direction the perimeter is independent of the
position. 

\section{Conclusions and discussion}

In the present paper  a six-dimensional low energy model hes been presented.
The model contains four-branes and empty bulk. It appears, that in a two four-brane 
case the  requirement of the usual four-dimensional Einstein gravity and non-existence
of unwanted light Kaluza-Klein modes constraint the model so, that the various
modes of gauge-singlet fermion fields decouple. The mass  spectrum
of these modes, given by (\ref{effmass}), contains only heavy effective fermions
(besides the zero mode), if the effective size of both directions of the torus are
small enough. In the present model it is possible to same time have large Kaluza-Klein
masses and keep the volume of the torus large enough, {\it i.e.} pull the gravitational 
scale $M_*$ down near to its experimantal limit.
Thus for each gauge-singlet fermion field there is only one light 
fermionic mode which couples to ordinary matter, {\it e.g.} to left-handed 
neutrinos. Indeed, mixing
with ordinary left-handed neutrinos on the brane(s) results hierarchy of neutrino
masses and mixings as studied by several authors.

The model has, of course, many open questions, including the stability of the extra 
dimensions, the mass spectrum of graviton excitations and the corrections to 
Newton's law, the neutrino masses and mixing on the brane. In the present paper,
we have introduced only ideas, how viable models can be possibly constructed on
a curved, compact manifold. More work is certainly needed to construct fully
realistic model.



\end{document}